\documentclass[
    ,final            
  ]
  {aipproc}

\layoutstyle{6x9}

\begin{document}

\title{Origin and evolution of galactic spin from looking at galaxy pairs}


\author{Bernardo Cervantes-Sodi}{
  address={Korea Astronomy and Space Science Institute, 61-1 Hwaam-dong, Yuseong-gu, Daejeon 305-348, Korea
   \footnote{bcsodi@kasi.re.kr}}
}

\author{X. Hernandez}{
  address={Instituto de Astronom\'\i a,
Universidad Nacional Aut\'onoma de M\'exico
A. P. 70--264,  M\'exico 04510 D.F., M\'exico}
}

\author{Changbom Park}{
  address={Korea Institute for Advanced Study, Dongdaemun-gu, Seoul 130-722, Korea}
}

\begin{abstract}
To study if the angular momentum gain for each member of a galaxy pair was
the result of tidal torques imprinted by the same tidal field, we search
for correlations between the spin in pairs of spiral galaxies identified
using the Sloan Digital Sky Survey. We find a weak, but statistically
significant correlation between the spin magnitude of neighbouring galaxies.
We show that events such as close interactions with neighbours play an
important role in the value of the spin for the final configuration, as
we find these interactions tend to reduce the value of the $\lambda$ spin
parameter of late-type galaxies considerably. This implies that the
original tidal field for each pair could have been similar, but the
redistribution of angular momentum at later stages of evolution is
important.
\end{abstract}

\maketitle


\section{Spin correlations}

In the simplest version of the tidal torque theory, both members of a galaxy
pair are immersed in the same tidal field, which, for similar galaxies, imprints the same amount of
angular momenta per unit mass to each member. In this context we would expect
a correlation between the spin mass product of galaxy pairs. In order to search for
such correlation we employed a sample of galaxy pairs selected by \cite{Park},
using the Sloan Digital Sky Survey, imposing the following criteria:
the neighbour galaxy can not be fainter than the target galaxy by more
than $\Delta M_{r}=0.5$, (2) it must have the smallest projected separation across
the line of sight from the target galaxy and (3), present a radial velocity
difference less than $V_{max}$, a value determined empirically (see \cite{Park}).
To account for the magnitude of the galactic angular momentum we employed
the $\lambda$ spin parameter, and using the simple model presented in
\cite{Hernandez:2006}, we calculate its value for each galaxy involved.
In Figure 1 left panel, is shown the  $\lambda_{1} M_{1}$ product of target
galaxies against  $\lambda_{2} M_{2}$ of nearest neighbour galaxies for the
pairs of the sample, presenting a low correlation with a correlation index of
$r^{2}=0.247$, but still statistically significant when comparing with the results
of a Monte Carlo test, assuming no intrinsic correlation, taking into account the observational constrains of the sample, yielding $r^{2}=0.129 \pm 0.063$.

\section{Galaxy interactions}

In the right panel of Figure 1 is ploted the value of $\lambda$ of each target galaxy
as a function of the distance to its nearest neighbour, normalized to the virial radius
of the neighbour galaxy, where the median $\lambda$ values are shown with their
dispersion presented as thin error bars and the uncertainty represented by thick error bars.
We can clearly
see the effects of galaxy-galaxy interaction, these appear as soon as the galaxies cross into their 
virial radii, leading to a gradual decrease in the values of $\lambda$. A complementary study
taking into account spin orientations and alignments is presented in \cite{Cervantes:2009}.

Our results do not diminishes the importance of the torquing at early stages of galaxy formation (indeed, the above random correlation of $\lambda_{1}M_{1}$ vs. $\lambda_{2}M_{2}$ is an empirical validation of the tidal torque theory), but give us an insight into the important role
of later interactions in the overall evolution of the total angular momentum of the galaxies, as shown
by the results obtained from interacting systems, where the spin is visibly perturbed by the presence of a
nearby companion.

\begin{figure}
\begin{tabular}{lr}
\includegraphics[height=.3\textheight]{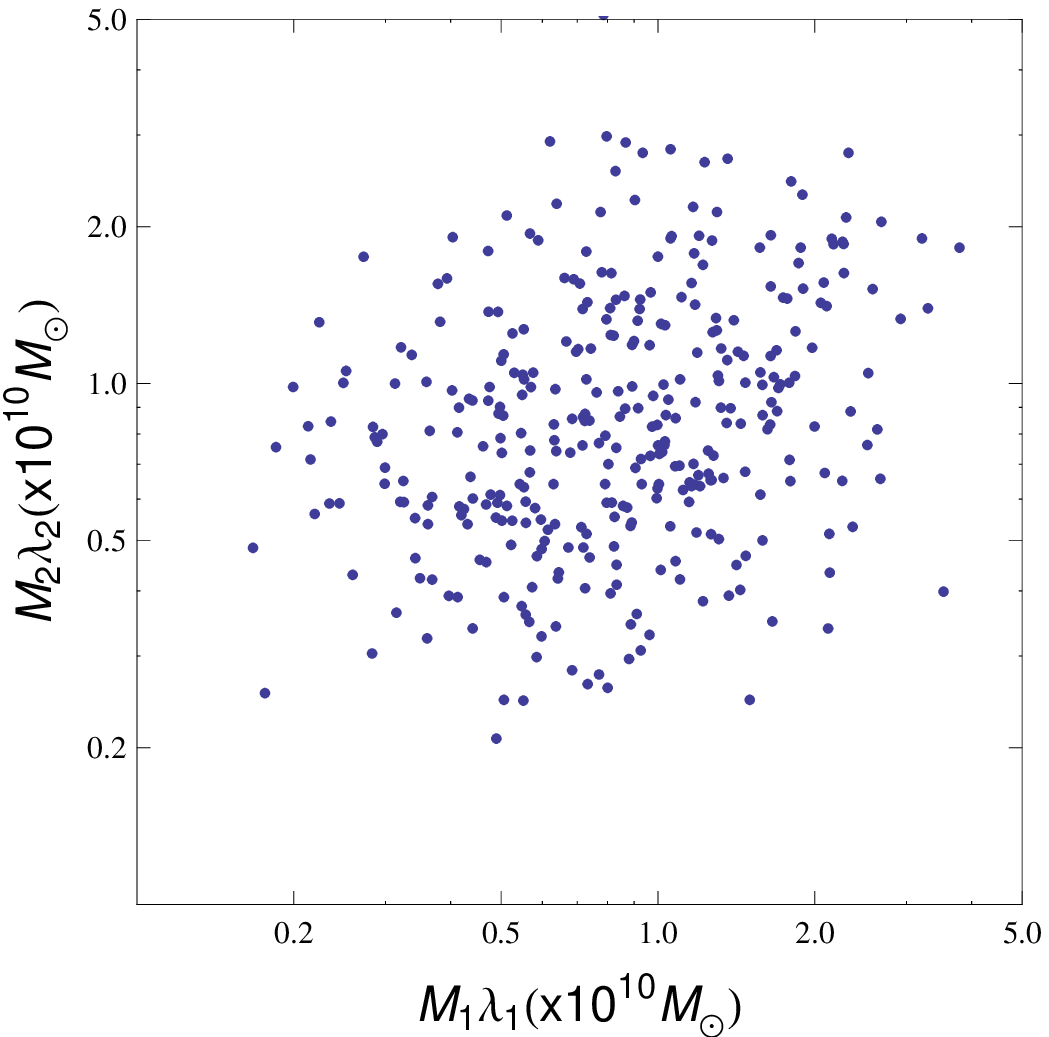} & \includegraphics[height=.3\textheight]{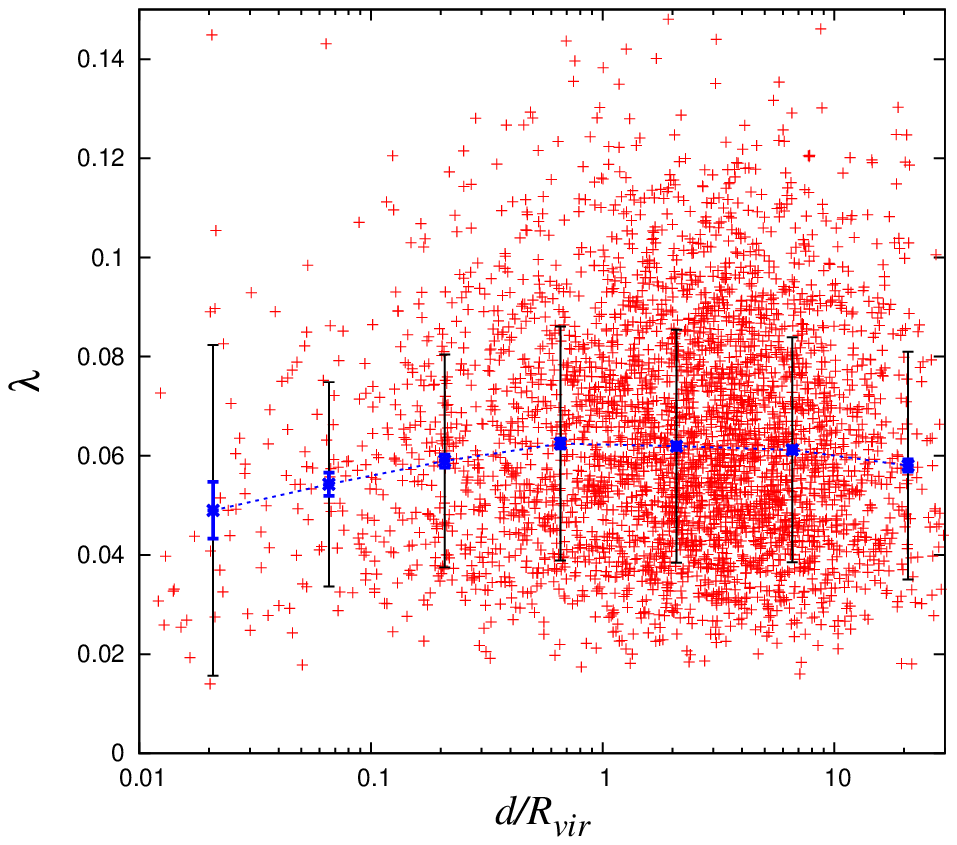} \\
\end{tabular}
\caption{\textit{Left panel:} $\lambda_{1} M_{1}$ product of target galaxies against  $\lambda_{2} M_{2}$ of nearest neighbour galaxies. \textit{Right panel:} $\lambda$ value for target late type galaxies as a function of the separation distance
to their nearest neighbour, normalized by the virial radius of the neighbour galaxy  }
\label{fig:1}
\end{figure}






\end{document}